\documentclass{article}

\usepackage[pdftex]{graphicx,color}
\usepackage{fullpage}
\usepackage{natbib}

\title{Contaminant remediation decision analysis using information gap theory}
\author{Dylan R. Harp (dharp@lanl.gov) and Velimir V. Vesselinov (vvv@lanl.gov)\\Computational Earth Sciences Group\\Los Alamos National Laboratory\\Los Alamos, NM, USA}

\begin{document}


\maketitle

\begin{abstract}
Decision making under severe lack of information is a ubiquitous situation in nearly every applied field of engineering, policy, and science. A severe lack of information precludes our ability to determine a frequency of occurrence of events or conditions that impact the decision; therefore, decision uncertainties due to a severe lack of information cannot be characterized probabilistically. To circumvent this problem, information gap (info-gap) theory has been developed to explicitly recognize and quantify the implications of information gaps in decision making. This paper presents a decision analysis based on info-gap theory developed for a contaminant remediation scenario. The analysis provides decision support in determining the fraction of contaminant mass to remove from the environment in the presence of a lack of information related to the contaminant mass flux into an aquifer. An info-gap uncertainty model is developed to characterize uncertainty due to a lack of information concerning the contaminant flux. The info-gap uncertainty model groups nested, convex sets of functions defining contaminant flux over time based on their level of deviation from a nominal contaminant flux. The nominal contaminant flux defines a reasonable contaminant flux over time based on existing information. A robustness function is derived to quantify the maximum level of deviation from nominal that still ensures compliance for each decision. An opportuneness function is derived to characterize the possibility of meeting a desired contaminant concentration level. The decision analysis evaluates how the robustness and opportuneness change as a function of time since remediation and as a function of the fraction of contaminant mass removed.
\end{abstract}

\section{Introduction}
\label{sect:Intro}
Environmental and earth scientists are frequently required to provide scientifically defensible support in decision-making processes related to important ecological problems (e.g.\ climate change, contaminant migration, carbon sequestration, nuclear waste storage, etc.\ \citep{Harrington85,Caselton92,Min05}). The decisions are often based on analyses of predictions obtained with system models representing the physical processes and conditions related to the problem. For example, hydrogeologists regularly provide modeling decision support to aid in the selection of contaminant remediation strategies \citep{Tartakovsky07}. In these cases, the model-based decision support is often driven by model predictions of contaminant concentrations at a point of regulatory compliance. However, uncertainties in the model predictions (predictive uncertainties) generally complicate the decision analysis. Predictive uncertainties result from limits in existing information (information is used here to refer to knowledge and data) about (1) governing processes, (2) boundary and initial conditions, and (3) state variables and process parameters. 

The status quo is to estimate probabilistic uncertainties in the physical process model inputs (prior uncertainties) and propagate these uncertainties through the physical process model to obtain estimates of predictive uncertainties. This approach is commonly utilized in Bayesian decision analysis (\citep{Schwede08}), and is a sound and justifiable approach when the uncertainty of each combination of model inputs and conditions can be characterized probabilistically or by a frequency of occurrence . However, most decisions related to environmental remediation frequently include uncertainties due to a severe lack of information, and cannot be characterized probabilistically or by frequency of occurrence. These types of uncertainties can be considered Knightian uncertainties, after the economist Frank Knight, who distinguished risk, which can be quantified in a lottery sense, and uncertainty, which, in his definition, is immeasurable \citep{Knight21}.

Therefore, in general, decision analyses providing the probabilistic confidence of success associated with a particular decision are unrealistic and unreliable as the probability distribution functions (pdf's) of the potential events are unknown. In spite of this limitation, estimates for the confidence of success or failure of decisions are commonly requested and provided for by these types of decision analyses, even when the assumptions required to obtain the probabilities of events are highly questionable \citep{Ben-Haim06}.

Probabilistic attempts to deal with a severe lack of information require invocation of the ''Principle of Indifference`` (i.e.\ an assumption in probability theory that all currently conceivable events are equally probable). This ``Principle'' is applied to justify the use of non-informative priors in Bayesian theory. However, the validity of this ``Principle'' in a decision analysis cannot be verified  \citep{Ben-Haim06}. 

A probabilistic analysis of uncertainties due to a lack of information are brought further into question if the concept of a ``collective'' advocated by, among others, \cite{vonMises39}, is taken in consideration. According to von Mises, probabilities are meaningless outside of a collective. For example, the probability that a 40-year-old man may die in the next year will be significantly different than the probability that a 40-year-old man who smokes will die in the next year, even though the same person can be a member of both collectives. Therefore, probabilities are only relevant within the context of a collective, and are meaningless when applied to a single element that can be grouped within multiple collectives. In cases of environmental remediation under severe lack of information, where the important processes and properties are characterized vaguely at best, it is hard to imagine an appropriate definition of a ``collective'', not to mention a dataset capable of characterizing the probability of success or failure for this ``collective''. Applying model-based Bayesian decision analyses under severe lack of information require a leap of faith in assuming that the collective is a set of predictions produced by system models whose ability to correctly represent all potential events cannot be verified due to the lack of information.

In general, environmental and earth scientists often encounter problems where the lack of information is so severe, that characterizing the probability of all possible events is infeasible. For example, contaminant concentration predictions may be highly dependent on infiltration events driven by precipitation and snowmelt, ultimately affecting the contaminant mass flux into an aquifer (infiltration is defined as a groundwater mass flux at the top of the regional aquifer here; infiltrated water originates on the ground surface and some of the groundwater carries the contaminant mass to the aquifer). Statistical characterization of infiltration events based on past records often provides poor predictions of the future probabilities of such events \citep{Wallis67,Kobold05}. The future predictions are additionally complicated when the predictive (compliance) period extends for a long period of time (for example, on the order of the millions of years in the case of nuclear waste repositories) which requires the consideration of the potential impact of climate changes (man-made and natural). For many natural phenomena, including infiltration, a strong potential exists to encounter a single extreme event or sequence of less extreme events outside what has been observed in the past. Uncertainties of this type are due to a gap in our information (Knightian uncertainties), and not an uncertainty related to which event in a set of events with known probabilities will occur. 

The need for non-probabilistic analyses of uncertainty in order to make reasonable environmental management decisions has been increasingly recognized. \cite{Hipel99} develop an info-gap decision analyses for water treatment facility design given a lack of information concerning the maximum possible flow rate. \cite{Levy00} combine multi-attribute value theory and info-gap decision theory to quantify the robustness of policy alternative to ecological info-gap uncertainties. \cite{Fox07} demonstrate an info-gap approach to calculate the power and sample size in ecological investigations with uncertain design parameters and distributional form. \cite{McCarthy07} develop an info-gap decision analysis to evaluate timber production and urban water supply management alternatives subjected to an info-gap uncertainty in fire risk. \cite{Stranlund08} developed an info-gap decision analysis to choose between price-based and quantity-based environmental regulation. \cite{Hine10} developed an info-gap decision analysis for flood management to account for info-gap uncertainties in flood models. \cite{Riegels11} evaluate the effects of info-gaps in hydro-economic model inputs on the selection of water price and target value for an ecological status parameter. In this paper, we develop an info-gap decision analysis on a contaminant remediation scenario where an info-gap exists concerning the contaminant mass flux into an aquifer. 

\section{Info-gap theory}
\label{sect:ig_theory}

The info-gap theory provides a general theoretical framework for decision analyses. An info-gap decision analysis for a specific problem requires three components: (1) model appropriately characterizing system behavior, (2) decision uncertainty model consistent with the info-gap theory, (3) decision performance goals (required and desired). These components are used to derive immunity functions, robustness and opportuneness functions, characterizing the immunity to failure and immunity to windfall success, respectively, of alternate decisions. 

\subsection{System model}
\label{sect:model}

The system model in an info-gap decision analysis characterizes the system performance based on the alternate decisions subjected to the ambient uncertainty. For environmental management decision scenarios, this will generally be a physical process model characterizing the natural and man-made processes controlling critical outputs influencing the decision.

\subsection{Info-gap uncertainty model}
\label{sect:ig_unc_model}

Info-gap uncertainty models rank an information gap by the uncertainty parameter $\alpha$.  The uncertainty model is comprised of nested sets of uncertain entities (i.e.\ parameters, functions, etc.\ which have info-gap uncertainties) ranked by the largest information gap that can be included in the set \citep{Ben-Haim06}. This approach is in sharp contrast to probabilistic or fuzzy logic approaches to uncertainty, which distribute uncertainty across all potential events to define recurrence-frequency or plausibility \citep{Ben-Haim06}. Info-gap uncertainty models provide less constraints and are intended for cases where lack of information precludes the ability to distribute uncertainties across all potential events, or even identify all potential events. Various types of info-gap decision uncertainty models include energy-bound, envelope-bound, slope-bound, and Fourier-bound models \citep{Ben-Haim06}. The selection and development of info-gap uncertainty models is scenario specific requiring few axiomatic constraints.

\subsection{Decision performance goals} 
\label{sect:dpg}

Performance goals in an info-gap decision analysis express a required or desired reward. In environmental decision scenarios, a required performance goal is commonly a constant fixed by a regulatory standard (e.g.\ maximum concentration limit; MCL). A desired performance goal is not a regulatory requirement, but may entail a more stringent goal than the regulatory standard. For instance, it may be desirable by decision makers or stakeholders to meet a stringent health standard that is below the regulatory standard.

\subsection{Immunity functions}
\label{sect:if}

The immunity functions define the immunity to failure (robustness) and immunity to windfall (opportuneness) of alternate decisions. The robustness function defines the maximum horizon of uncertainty ($\alpha$) where failure cannot occur. As we typically lack the information to know the actual horizon of uncertainty, the info-gap uncertainty model is an unbounded function of the horizon of uncertainty in general. This can be expressed linguistically as 

\begin{equation}
	\widehat{\alpha}(q) = \max\{\alpha:\ \mbox{the required performance goal is satisfied}\}
	\label{eq:alpha}
\end{equation}

\noindent where $q$ is a vector containing the alternate decisions and $\widehat{\alpha}(q)$ is the robustness function. 

The opportuneness function defines the minimum horizon of uncertainty ($\alpha$) where windfall success cannot occur. Large values of opportuneness indicate that large deviations from nominal (large ambient uncertainty) are needed in order to enable the potential of exceptional success. Small values of opportuneness indicate that a low ambient uncertainty provides the potential for exceptional success. Linguistically, opportuneness can be expressed as

\begin{equation}
	\widehat{\beta}(q) = \min\{\alpha :\ \mbox{the possibility of meeting the desired performance goal exists}\}
	\label{eq:beta}
\end{equation}

\noindent where $\widehat{\beta}(q)$ is the opportuneness function. 

The complimentary nature of robustness and opportuneness are evident. The robustness and opportuneness can be sympathetic or antagonistic in a decision analysis, depending on the particular scenario.

\section{Contaminant remediation decision scenario}
\label{sect:crds}

\begin{figure}
\begin{center}
	\includegraphics[width=8cm]{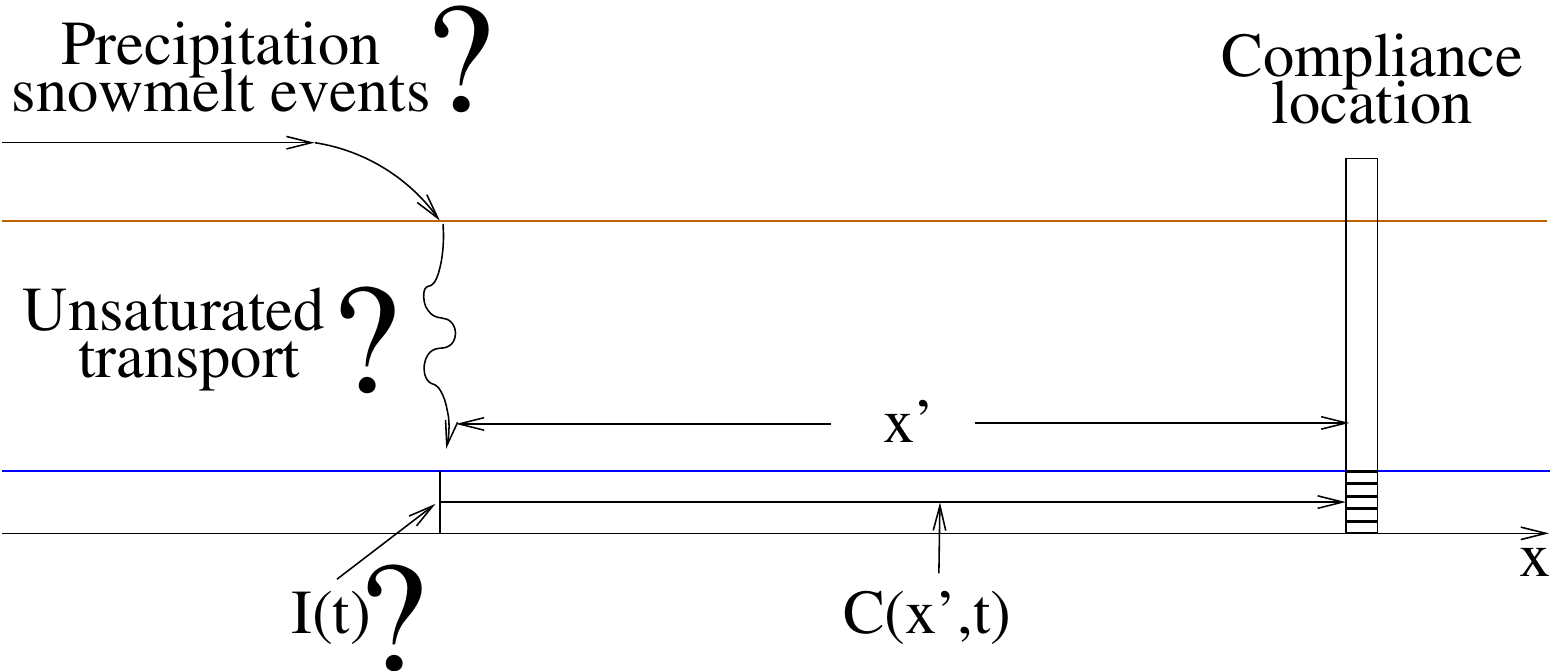}
	\caption{Contamination remediation scenario diagram}
	\label{fig:dia}
\end{center}
\end{figure}

The decision scenario of contaminant remediation presented below is representative of an actual case study at Los Alamos National Laboratory (LANL) related to an existing contamination site. A diagram of the contaminant spill scenario is presented in Figure~\ref{fig:dia} and described below. A contaminant spill with known mass has been released on the ground surface and is spatially distributed in the soil below the release location. The contaminant is known to chemically degrade over time (for example, due to radioactive decay or chemical hydrolysis). An aquifer utilized for municipal water supply lies below the contaminated soil. A compliance point is located near the spill where regulatory health standards dictate the maximum contaminant concentration. Exceeding the regulatory standard will compromise the municipal water supply, incur fines from the regulatory agency, and compromise the integrity of those involved in the remediation effort. Removing the contaminant from the soil is expensive, and entails risks of exposure to workers and redistribution of the contaminant in the environment. A decision analysis is desired to determine the robustness of selecting various fractions of the original mass to remove in order to ensure regulatory compliance given an info-gap in the contaminant mass flux into the aquifer (contaminant plume source strength). The proposed info-gap analysis can be applied to physical process models with different complexity. The analysis presented below uses a relatively simple analytical model, which can be considered a first step in a tiered process that utilizes more complicated models in subsequent stages.

\subsection{Contaminant transport model}
\label{ctm}

An analytical solution describing the two-dimensional advective-dispersive transport of a contaminant within an aquifer is \citep{Wang09}

\begin{eqnarray}
\nonumber	C(x,y,t) &=& \frac{1}{4 \pi n \sqrt{D_xD_y}} \int_0^t I(t - \tau) \exp\left [ -\lambda \tau - \frac{(x-u\tau)^2}{4D_x\tau} - \frac{y^2}{4D_y\tau} \right ] \frac{\mathrm{d}\tau}{\tau},\\
	 &&-\infty < x,y < \infty, t>0,
\end{eqnarray}

\noindent where $C(x,y,t)$ [ML$^3$] is a contaminant concentration in the aquifer, $I(t)$ [ML$^{-1}$T$^{-1}$] is the transient contaminant flux (source strength) at the point $x=y=0$ per unit depth of the aquifer, $n$ is the porosity, $D_x$ and $D_y$ are the principal dispersion coefficients [L$^2$T$^{-1}$], $\lambda$ [T$^{-1}$] is the first-order constant of decay, and $u$ [LT$^{-1}$] is the pore water velocity. The groundwater flow is along the $x$-direction. Assuming that the point of compliance is located directly downgradient from the plume source along the $x$ axis, we can simplify the model by setting $y=0$ as

\begin{equation}
	C(x,t) = \frac{1}{4 \pi n \sqrt{D_xD_y}} \int_0^t I(t - \tau) \exp\left [ -\lambda \tau - \frac{(x-u\tau)^2}{4D_x\tau} \right ] \frac{\mathrm{d}\tau}{\tau}.
	\label{eq:C}
\end{equation}

\noindent Let us define an impulse response function,

\begin{equation}
	h(x,t)= \frac{1}{4 \pi n t \sqrt{D_x D_y}} \exp \left [-\lambda t - \frac{(x-ut)^2}{4D_xt} \right ],
	\label{eq:h}
\end{equation}

\noindent and substitute this into equation~\ref{eq:C} allowing the system model that will be applied in the info-gap analysis to be defined as

\begin{equation}
	C(x,t)=\displaystyle \int_0^t I(t - \tau)h(x,\tau) \mathrm{d}\tau.
	\label{eq:sys}
\end{equation}

The functional form of equation~\ref{eq:sys} can be used in general to describe the effect of an impulse on a system. Therefore, while the application presented here is contaminant remediation with uncertain contaminant flux, much of the development of the decision analysis presented here can be applied to other decision analyses with analogous uncertainties due to unknown impulse functions.

\subsection{Contaminant flux info-gap uncertainty model}
\label{sect:cfigum}

The info-gap uncertainty model for the contaminant flux into an aquifer is defined as the potential for deviations in the contaminant flux from a nominal value, and can be expressed as

\begin{equation}
	\mathcal{U}(\alpha,\widetilde{I} (t) ) = \left \{ I(t):  \frac{ \int_0 ^{t} [I(t) - \widetilde{I}(t) ]^2 \mathrm{d}t }{\int_0 ^{t} \widetilde{I}^2(t) \mathrm{d}t } \leq \alpha^2 \right \},\ \alpha \geq 0,
	\label{eq:U}
\end{equation}

\noindent where $\widetilde{I}(t)$ is the nominal contaminant flux function and $\alpha$ defines levels of info-gap uncertainty describing deviation of the contaminant flux from nominal. Equation~\ref{eq:U} defines an info-gap uncertainty model representing nested, convex sets of contaminant flux functions $I(t)$. Functions in these sets can contain a single extremely large event, a high frequency of relatively smaller events, or any combination thereof, as long as $\int_0 ^{t} [I(t) - \widetilde{I}(t) ]^2 \mathrm{d}t / \int_0^t \widetilde{I}^2(t) \mathrm{d}t \leq \alpha^2$. 

Equation~\ref{eq:U} presents an instance of an energy-bound info-gap uncertainty model. Energy-bound models have the ability to capture uncertainties in transients, where prior information concerning the potential for large deviations or series of small deviation is extremely limited.

\subsection{Contaminant concentration performance goals} 
\label{sect:perform}
In the current scenario, the required performance goal is fixed by regulatory standards. The performance requirement is defined as the regulatory limit on the contaminant concentration (e.g. MCL) at the point of compliance as 

\begin{equation}
	C(x',t) \leq C_{c},\ \forall\ t>0, 
	\label{eq:perf}
\end{equation}

\noindent where $x'$ is a point of compliance (e.g.\ site boundary, pumping well) and $C_c$ is the critical contaminant concentration based on a regulatory standard.

In decision analyses, frequently, there is a desired performance goal that is not strictly required but would be beneficial if met. This allows us to explore the opportunity of achieving this performance given alternative decisions. In a contaminant remediation decision scenario, the desired system performance may be a recommended contaminant concentration threshold that is less than the regulatory standard. The desired performance goal is described as

\begin{equation}
	C(x',t) \leq C_{w},\ \forall\ t>0,
	\label{eq:wf}
\end{equation}

\noindent where $C_w$ is the desired contaminant concentration.

The performance goals expressed in inequalities~\ref{eq:perf} and \ref{eq:wf} illustrate the fact that uncertainty can be both pernicious, causing failure, and propitious, enabling the potential of exceptional windfall success  \citep{Ben-Haim06}. For example, the ambient uncertainty is pernicious when making a decision to ensure the performance requirement of inequality~\ref{eq:perf}, while it is propitious when making a decision to allow the potential to surpass the performance expressed in inequality~\ref{eq:wf}.

\subsection{Robustness function}
\label{sect:rob}
Considering the contaminant flux uncertainty model (equation~\ref{eq:U}) and the performance requirement (equation~\ref{eq:perf}), the decision robustness function can be expressed as

\begin{equation}
	\widehat{\alpha}(q,C_c) = \max \left \{ \alpha :\ \left ( \max\limits_{I \in \mathcal{U}(\alpha, \widetilde{I})} C(x',t,q) \right ) \leq C_c \right \},\ \forall\ t > 0,
	\label{eq:alpha0}
\end{equation}

\noindent where $q$ is the fractional percent of the contaminant mass removed, defined as $q=M_r/M_t$, where $M_r$ is the mass removed at $t=0$ and $M_t$ is the total mass released in the environment. The robustness function $\widehat{\alpha}$ is dimensionless. More complicated schedules for contaminant removal can also be applied: for example, mass removal within a given period of time, or periodically over several periods. The contaminant flux into the aquifer and contaminant concentrations in the aquifer will decrease with increasing $q$, therefore $I=f(t,q)$ and $C=f(x,t,q)$. 

Equation~\ref{eq:sys} can be expressed as the addition of the nominal concentration and the deviation from the nominal concentration at location $x'$ as

\begin{equation}
	C(x',t,q) = \underbrace{\int_0^{t} \widetilde{I}(t - \tau,q) h(x',\tau) \mathrm{d}\tau}_{\widetilde{C}(x',t,q)} + \underbrace{\int_0^{t} [I(t - \tau,q)-\widetilde{I}(t - \tau,q)] h(x',\tau) \mathrm{d}\tau}_{C(x',t,q)-\widetilde{C}(x',t,q)}
	\label{eq:split}
\end{equation}

\noindent where the nominal concentration $\widetilde{C}(x',t,q)$ is the concentration resulting from the nominal contaminant flux. An upper limit can be determined for the second integral in equation~\ref{eq:split} by using the Schwarz inequality \citep{Weisstein} as

\begin{equation}
	\left ( \int_0^{t} [I(t - \tau,q)-\widetilde{I}(t - \tau,q)] h(x',\tau) \mathrm{d}\tau \right )^2 \leq \int_0^{t} [I(\tau,q)-\widetilde{I}(\tau,q)]^2 \mathrm{d}\tau \int_0^{t} h(x',\tau)^2 \mathrm{d}\tau
	\label{eq:schwarz}
\end{equation}

\noindent Using inequality~\ref{eq:schwarz} in equation~\ref{eq:split} leads to the following inequality:

\begin{equation}
	C(x',t,q) \leq  \widetilde{C}(x',t,q) + \sqrt{\int_0^{t} [I(\tau,q)-\widetilde{I}(\tau,q)]^2 \mathrm{d}\tau \int_0^{t} h(x',\tau)^2 \mathrm{d}\tau}
	\label{eq:ineq}
\end{equation}

\noindent Considering the info-gap uncertainty model (equation~\ref{eq:U}), it is recognized that

\begin{equation}
	\int_0^{t} [I(\tau,q)-\widetilde{I}(\tau,q)]^2 \mathrm{d}\tau \leq \alpha^2 \int_0^t \widetilde{I}^2(\tau) \mathrm{d}\tau.
\end{equation}

\noindent Substituting this into inequality~\ref{eq:ineq}, a maximum concentration at $x'$ up to uncertainty $\alpha$ can be defined as

\begin{equation}
	\max\limits_{I \in \mathcal{U}(\alpha,\widetilde{I})} C(x',t,q) = \widetilde{C}(x',t,q) + \alpha \sqrt{ \int_0^t I^2(\tau,q) \mathrm{d}\tau \int_0^{t} h^2(x',\tau) \mathrm{d}\tau}
	\label{eq:max}
\end{equation}

\noindent Setting the maximum concentration equal to $C_c$, as defined by inequality~\ref{eq:perf}, and solving for $\alpha$ results in the robustness function as a function of time

\begin{equation}
	\widehat{\alpha}(q,t) = \frac{C_c - \widetilde{C}(x',t,q)}{\sqrt{\int_0^t I^2(\tau,q) \mathrm{d}\tau \int_0^{t} h^2(x',\tau) \mathrm{d}\tau}}.
	\label{eq:alphahat}
\end{equation}

\noindent As inequality~\ref{eq:perf} requires compliance at all times, the robustness function considering all times is

\begin{equation}
	\widehat{\alpha}(q) = \min\limits_{t>0}\ \widehat{\alpha}(q,t).
	\label{eq:minalpha}
\end{equation}

\noindent where the robustness is dimensionless and defines the maximum fractional error in the actual contaminant flux from nominal that still ensures compliance for alternative decisions.

\subsection{Opportuneness function}
\label{sect:of}
Considering the desired performance described by equation~\ref{eq:wf}, a complimentary equation to equation~\ref{eq:alpha0} can be defined for the opportuneness function $\widehat{\beta}$, also dimensionless, as 

\begin{equation}
	\widehat{\beta}(q,C_w) = \min \left \{ \alpha :\ \left ( \min\limits_{I \in \mathcal{U}(\alpha, \widetilde{I})} C(x',t,q) \right ) \leq C_w \right \},\ \forall\ t > 0.
	\label{eq:beta0}
\end{equation}

\noindent The complimentary nature of robustness and opportuneness is apparent by comparison of equations~\ref{eq:alpha0} and \ref{eq:beta0}.

In our decision scenario, the opportuneness function quantifies the least level of uncertainty required to maintain the potential that $C(x',t,q)$ will not exceed $C_w$. This leads to an equation complimentary to equation~\ref{eq:max} defining the minimum possible concentration up to uncertainty $\alpha$ as

\begin{equation}
	\min\limits_{I \in \mathcal{U}(\alpha,\widetilde{I})} C(x',t,q) = \widetilde{C}(x',t,q) - \alpha \sqrt{ \int_0^t I^2(\tau,q) \mathrm{d}\tau \int_0^{t} h^2(x',\tau) \mathrm{d}\tau}
	\label{eq:min}
\end{equation}

Setting the minimum concentration equal to $C_w$ and solving for $\alpha$ produces the opportuneness function (complimentary to the robustness function; equation~\ref{eq:alphahat}) as

\begin{equation}
	\widehat{\beta}(q,t) = \frac{\widetilde{C}(x',t,q) - C_w}{\sqrt{\int_0^t I^2(\tau,q) \mathrm{d}\tau \int_0^{t} h^2(x',\tau) \mathrm{d}\tau}}.
	\label{eq:betahat}
\end{equation}

\noindent As the performance expressed by inequality~\ref{eq:wf} is desired at all times, the opportuneness function considering all times is

\begin{equation}
	\widehat{\beta}(q) = \max\limits_{t>0}\ \widehat{\beta}(q,t).
	\label{eq:maxbeta}
\end{equation}

\noindent where the opportuneness is dimensionless and defines the minimum fractional error in the actual contaminant flux from the nominal that maintains the possibility of meeting the desired performance goal. 

\section{Contaminant remediation info-gap decision analysis}
\label{crigda}

\begin{figure}
\begin{center}
	\includegraphics[width=7cm]{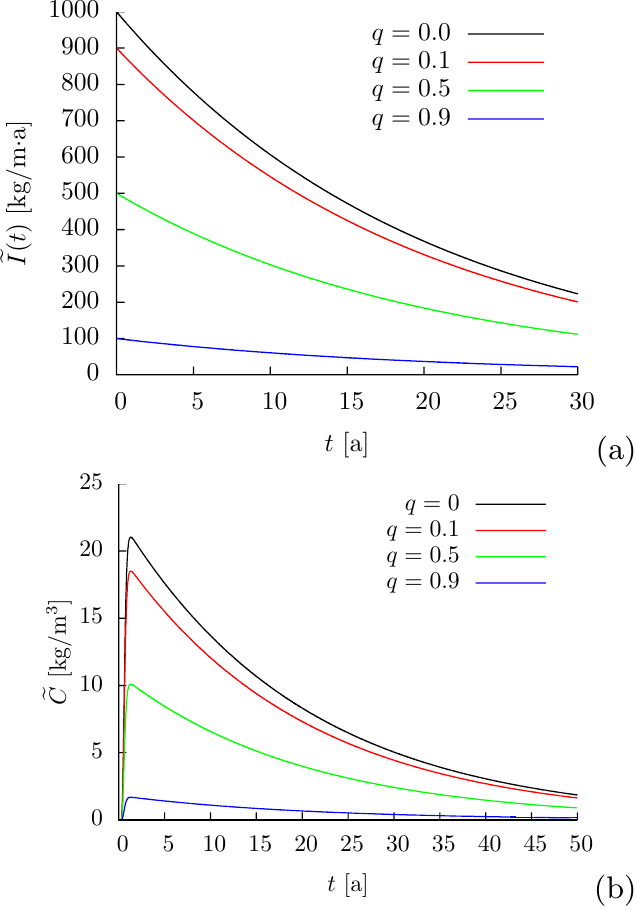}
	\caption{Nominal contaminant flux into the aquifer (a) and nominal contaminant concentration at the compliance point (b) over time since remediation for various fractions of contaminant removed $q$}
	\label{fig:I}
\end{center}
\end{figure}

The nominal contaminant flux into the aquifer is defined as $\widetilde{I}(t,q) = 1000*(1-q)*\exp[-0.05*t]$ [kg/m/a] and plotted for fractions of contaminant removed $q$ as a function of time since remediation in Figure~\ref{fig:I} (a). Constraining this estimate is not possible without further field studies or data acquisition. The following info-gap decision analysis will evaluate how wrong can our estimate of the contaminant flux into the aquifer be and still ensure compliance at various fractions of contaminant mass removal. The associated nominal predictions of concentration at the compliance point $x'=20$~m are plotted in Figure~\ref{fig:I} (b). It is assumed that the critical regulated concentration at $x'$ is $C_c=25$~kg/m$^3$ and that it would be a desirable outcome if the concentration did not exceed $C_w=5$~kg/m$^3$.

It is assumed that the parameters of the contaminant transport model (equation \ref{eq:C}) are well known and with negligible uncertainty compared to the info-gap in the contaminant flux. These parameters are defined as $D_x=30$~m$^2$/a, $D_y=7$~m$^2$/a, $n=0.1$, $\lambda=1$~/a, and $u=30$~m/a. These values are representative of the flow conditions at the LANL site. Extension of the current analysis by incorporation of probabilistic and info-gap uncertainties of these parameters is possible \citep{Hipel99,Ben-Haim06}. 

\begin{figure}
\begin{center}
	\includegraphics[width=9cm]{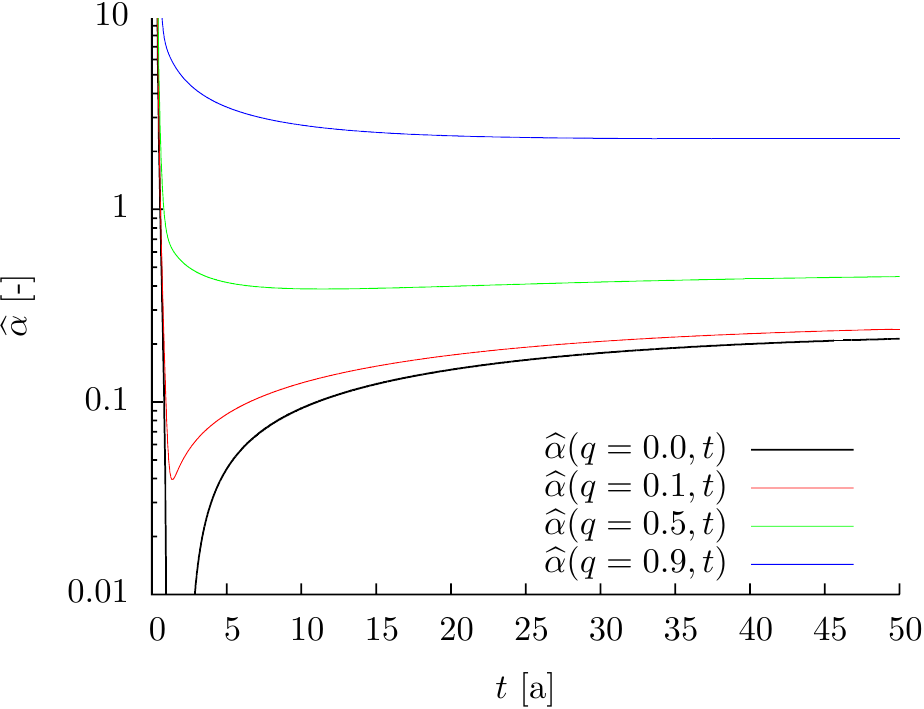}
	\caption{Robustness function versus time since remediation for various fractions of contaminant mass removed $q$. Note that robustness is plotted on a log scale.}
	\label{fig:alphavt}
\end{center}
\end{figure}

\begin{figure}
\begin{center}
	\includegraphics[width=9cm]{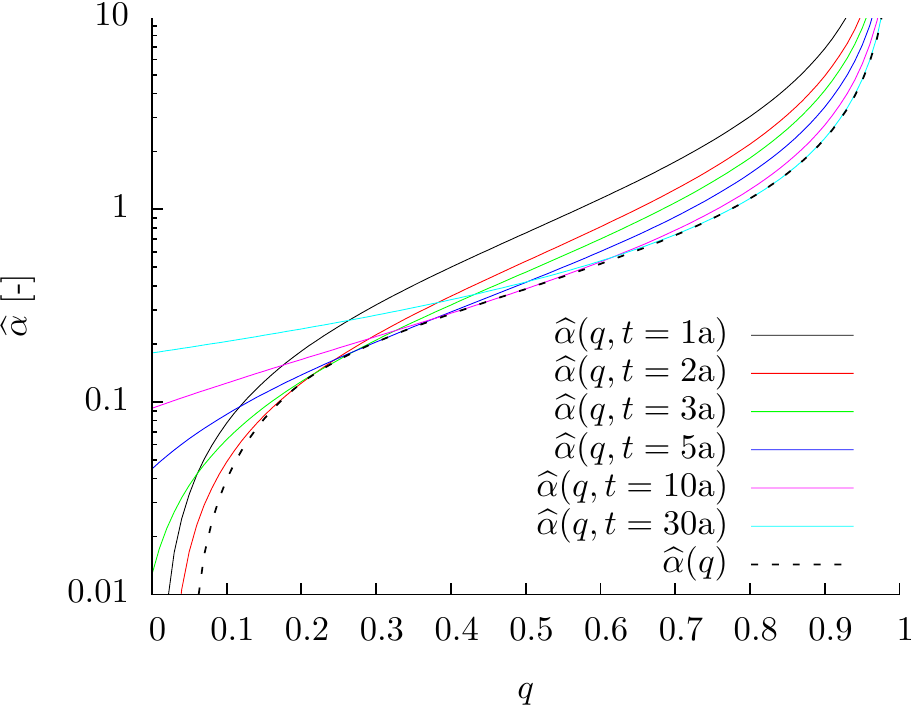}
	\caption{Robustness function versus the fraction of contaminant removed for various times since remediation (refer to equations~\protect\ref{eq:alphahat}); the dotted line plots the minimum robustness at any time (refer to equation~\ref{eq:minalpha}). Note that robustness is plotted on a log scale.}
	\label{fig:alphavq}
\end{center}
\end{figure}

\begin{figure}
\begin{center}
	\includegraphics[width=9cm]{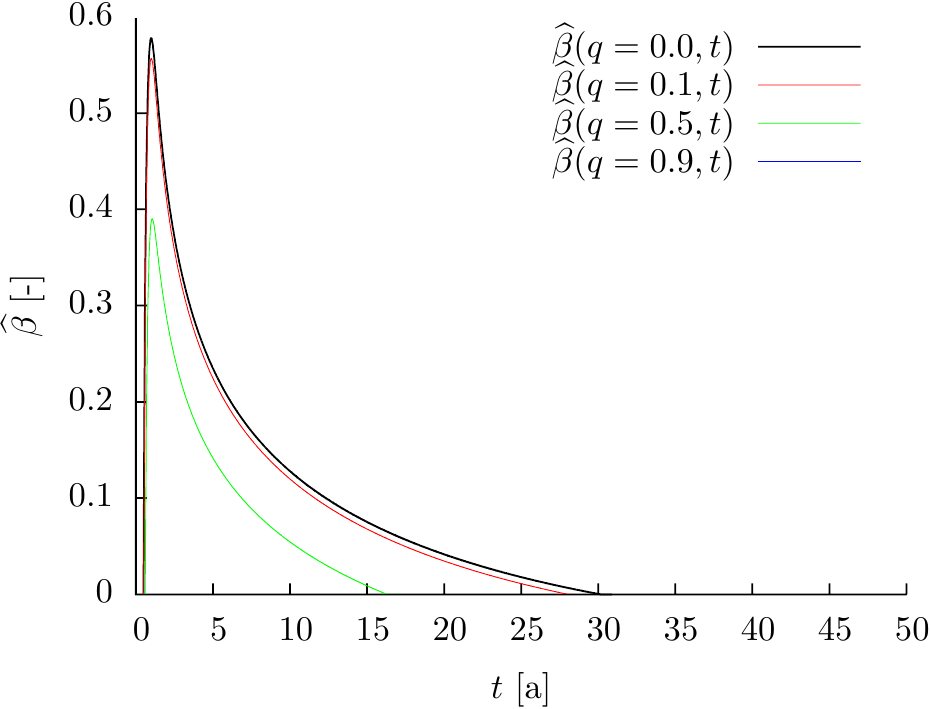}
	\caption{Opportuneness function versus time since remediation for various fractions of contaminant mass removed $q$. Note that $\widehat{\beta}(q=0.9,t)=0$ for all time.}
	\label{fig:betavt}
\end{center}
\end{figure}

\begin{figure}
\begin{center}
	\includegraphics[width=9cm]{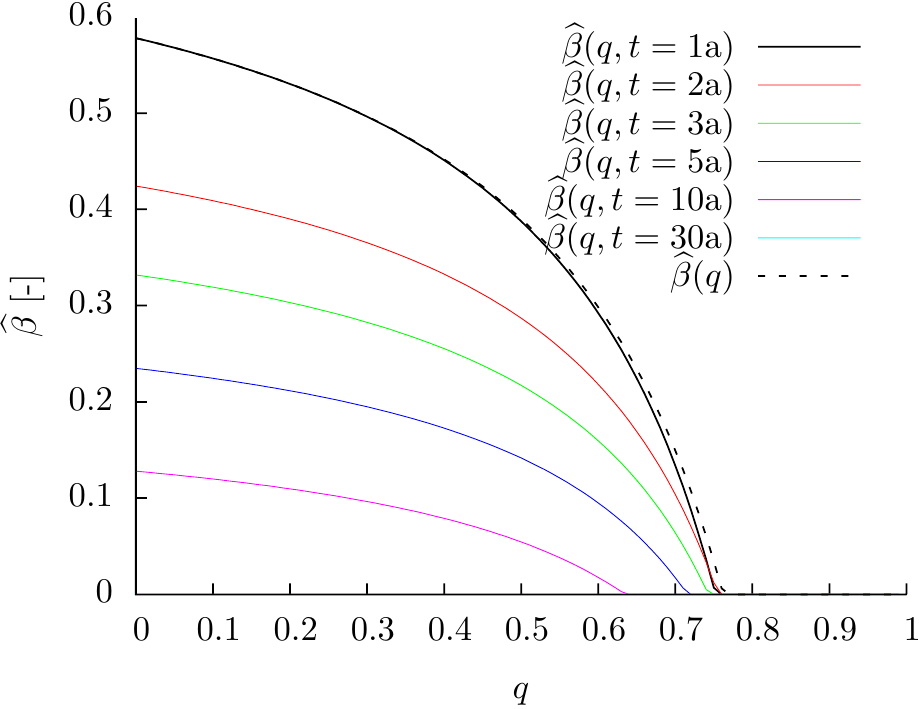}
	\caption{Opportuneness function versus the fraction of contaminant removed for various times since remediation (refer to equation~\protect\ref{eq:betahat}); the dotted line plots the maximum opportuneness for all times (refer to equation~\protect\ref{eq:maxbeta}). Note that $\widehat{\beta}(q,t=30a)=0$ for all $q$.}
	\label{fig:betavq}
\end{center}
\end{figure}

\begin{figure}
\begin{center}
	\includegraphics[width=9cm]{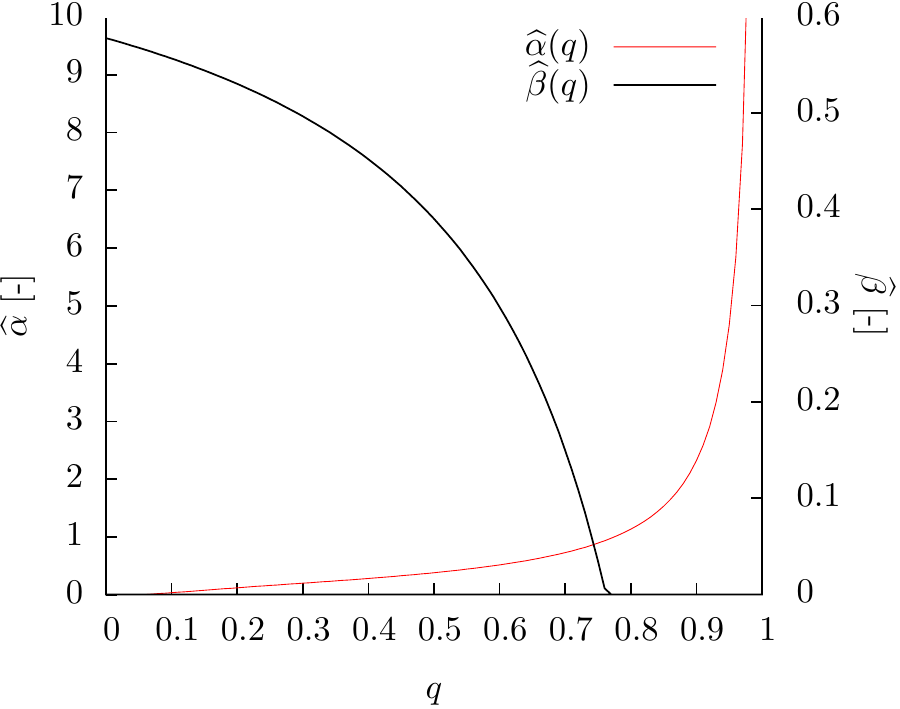}
	\caption{Robustness and opportuneness functions considering all times (refer to equations~\protect\ref{eq:minalpha} and \protect\ref{eq:maxbeta}, respectively) versus the fraction of contaminant removed. Note that both robustness and opportuneness are plotted on arithmetic scales in this figure.}
	\label{fig:rob_opp}
\end{center}
\end{figure}

The robustness function is plotted versus time since remediation for various fractions of contaminant mass removed $q$ in figure~\ref{fig:alphavt}. Robustness functions at a particular time since remediation versus the fraction of contaminant removed are plotted in figure~\ref{fig:alphavq} (refer to equation~\ref{eq:alphahat}. As we are interested in compliance at all times, the minimum robustness for each decision $q$ is also plotted as a dotted line in figure~\ref{fig:alphavq} (refer to equation~\ref{eq:minalpha}). In our example, robustness represents the maximum fractional error in the nominal contaminant flux that ensures that $C(x',t,q)<C_c$ (equation~\ref{eq:alpha0}). For example, a value of $\widehat{\alpha}=1$ indicates that the fractional error in the nominal can be 100\% (i.e.\ potential deviations from the nominal contaminant flux can be as high as twice the nominal contaminant flux), and the associated decision still ensures compliance.

Plots of the opportuneness functions are presented in figures~\ref{fig:betavt} and \ref{fig:betavq}. In this example, the opportuneness function represents the minimum fractional error in the nominal contaminant flux that sustains the possibility that $C(x',t,q)<C_w$ (equation~\ref{eq:beta0}). For example, a value of $\widehat{\beta}=0.1$ indicates that the relative error in the nominal contaminant flux must be at least 10\% to enable the possibility that the concentration will remain below the desired performance goal.

In figures~\ref{fig:alphavt}, \ref{fig:alphavq}, \ref{fig:betavt}, and \ref{fig:betavq}, the relationship between robustness/opportuneness and effort is apparent. Increased robustness and decreased opportuneness is only possible with increased effort and cost (proportional to the fraction of contaminant mass removed $q$). Interesting variations in robustness for given times since remediation as a function of $q$ are observed in figure~\ref{fig:alphavq}, demonstrating that for small values of $q$, late times have greater decision robustness, while for larger values of $q$, early times demonstrate greater robustness. Robustness for all times approach infinity as $q\rightarrow 1$, as removing all the contaminant will provide infinite robustness (of course, at a potentially unjustifiable cost). In figures~\ref{fig:betavt} and \ref{fig:betavq}, it is clear that the opportuneness increases ($\widehat{\beta}$ decreases) with time and fraction of mass removed. In figures~\ref{fig:betavt} and \ref{fig:betavq}, it can be determined that after 30 years, the opportuneness becomes zero for the decision to do nothing ($q=0$), while at 90\% removal ($q=0.9$), no uncertainty is necessary to allow the possibility that $C(x',t)<C_w$ for all times.

Figure~\ref{fig:rob_opp} plots the decision robustness (\ref{eq:alphahat}) and opportuneness (\ref{eq:betahat}) functions together. From equations~\ref{eq:alphahat} and \ref{eq:betahat}, the following expression can be derived to illustrate the complimentary relationship between robustness and opportuneness in the current decision scenario:

\begin{equation}
	\widehat{\beta}(q,t) = \frac{C_c-C_w}{\sqrt{\int_0^t I^2(\tau,q) \mathrm{d}\tau \int_0^{t} h^2(x',\tau)}} - \widehat{\alpha}(q,t),
\end{equation}

\noindent where it is apparent that as $\widehat{\alpha}$ increases, $\widehat{\beta}$ decreases. As it is desirable to select an alternative that increases $\widehat{\alpha}$ and decreases $\widehat{\beta}$, these two objectives are sympathetic in this decision scenario. An increase in robustness increases the opportuneness.

Figure~\ref{fig:rob_opp} can be used by a decision maker to evaluate the implications of the ambient uncertainty on alternative decisions. For example, at values of $q$ less than around 0.04 (removal of 4\% of the contaminant mass), the decision robustness is zero, indicating that failure to meet compliance (the required performance goal) is ensured based on the nominal contaminant flux. It should be noted that if the actual contaminant flux is lower than the nominal estimate, failure may not occur for low values of fraction removed. Decisions in this range also require the largest potential deviations from the nominal to enable the possibility of meeting the desired goal (equation~\ref{eq:wf})  (relative error in the contaminant flux of at least 57\%, or $\widehat{\beta}=0.57$, at $q=0.04$). A decision to remove approximately 6\% ($q=0.06$) of the contaminant mass will ensure compliance (the required performance goal) only if the actual contaminant flux deviates from the nominal by less than 1\% ($\widehat{\alpha}=0.01$), while the corresponding potential for exceptional success will require deviations from the nominal of at least approximately 56\% ($\widehat{\beta}=0.56$). Deciding to remove over approximately 76\% ($q=0.76$) of the mass ensures meeting the desired goal at zero deviation from the nominal (decisions in this range ensure that the concentration will be below $C_w$ based on the nominal contaminant flux), while compliance is ensured in this range at increasing potential deviation from nominal.  Deciding to remove 50\% ($q=0.5$) of the mass will ensure compliance if the actual contaminant flux deviates from the nominal by less than around 39\% ($\widehat{\alpha}=0.39$), while the corresponding potential for exceptional success will require the actual contaminant flux to deviate by at least 39\% ($\widehat{\beta}=0.39$). Other decisions can be evaluated similarly.

Based on Figure~\ref{fig:rob_opp}, the decision makers may want to select a decision in a range where (1) the robustness is greater than zero and (2) the opportuneness is greater than zero if there is relatively higher acceptance of potential risk (i.e.\ fraction of mass removal $q$ between $0.1$ and $0.75$). If decision makers prefer to select a decision with relatively lower risk, an alternative decision in the range where the opportuneness is equal or very close to zero ($q>0.75$) will provide higher immunity to failure. Decisions in the range where the robustness is equal or very close to zero ($q<0.1$) provide very low immunity to failure, and are potentially unacceptable.

In an actual application, there may be some concept of the cost associated with each $q$. The relationship between the cost and $q$ is not expected to be linear; typically, the cost increases sharply with the increase of $q$. As a result, analogous figures to figures~\ref{fig:alphavt} and \ref{fig:alphavq} can be formulated plotting decision robustness versus cost. A decision maker can use these plots to determine the cost to achieve different levels of robustness and opportuneness. This info-gap decision analysis can be extended to incorporate other info-gap or probabilistic uncertainties due to severe lack of information of other model inputs or conditions; for example, information regarding the groundwater velocity or the aquifer dispersion in the zone between the plume source and the compliance point can be extremely limited. 

\section{Conclusions}
\label{sect:concs}
Geoscientists are often confronted with decision scenarios related to environmental management where the lack of information precludes the ability to reasonably estimate probabilistic uncertainty models. In these cases, it is not possible to evaluate robustness in the context of the probability of exceeding a contaminant concentration at a compliance point \citep{Caselton92}. This paper demonstrates an approach that can be applied in these cases of severe uncertainty using an info-gap decision analysis. The proposed decision making framework can be applied for environmental management of contaminant remediation but also to problems such as radioactive waste storage, carbon sequestration, and climate change.

\section*{Acknowledgements}
This research was funded by the Environmental Programs Directorate of the Los Alamos National Laboratory. The authors extend their gratitude to Phoolendra K. Mishra, Greg M. Chavez, James R. Langenbrunner, Yakov Ben-Haim, Andy Wolfsberg, and Kay Birdsell for constructive discussions during the development of this paper.

\bibliographystyle{authordate1}
\bibliography{ref}

\end{document}